\documentclass[twocolumn]{aastex63}

\usepackage{times,color,natbib,url,amsmath}
\usepackage{graphicx,float,psfrag}
\usepackage{grffile}
\usepackage{tabularx}
\usepackage{latexsym,amsmath,amssymb}
\usepackage{natbib}
\usepackage{verbatim}
\usepackage{multirow}
\usepackage{color}
\usepackage[hang]{footmisc}
\usepackage{soul,xcolor}

\bibpunct{(}{)}{;}{a}{}{,}

\newcommand {\asca} {{\it ASCA}}

\newcommand {\swift} {\textsl{Swift}}
\newcommand {\rxte} {\textsl{RXTE}}
\newcommand {\fermi} {\textsl{Fermi}}
\newcommand {\nustar} {\textsl{NuSTAR}}

\def \rsun {\ifmmode$R$_{\odot}\else R$_{\odot}$}

\def \hcm {\hbox {\ifmmode $ atoms cm$^{-2}\else atoms cm$^{-2}$\fi}}

\def\approxgt{\mathrel{\hbox{\rlap{\lower.55ex \hbox {$\sim$}}
        \kern-.3em \raise.4ex \hbox{$>$}}}}
\def\approxlt{\mathrel{\hbox{\rlap{\lower.55ex \hbox {$\sim$}}
        \kern-.3em \raise.4ex \hbox{$<$}}}}

\def \arcsec {\hbox{$^{\prime\prime}$}}

\newcommand\T{\rule{0pt}{2.6ex}}
\newcommand\B{\rule[-1.2ex]{0pt}{0pt}}

\def \src {1RXS~J1708$-$40}

\begin{document}
\setstcolor{red}

\title{\rm \uppercase{Simultaneous magnetic polar cap heating during a flaring episode from the magnetar \\1RXS J170849.0-400910}}

%
\author{George~Younes}
\affiliation{Department of Physics, The George Washington University, Washington, DC 20052, USA, gyounes@gwu.edu}
\affiliation{Astronomy, Physics and Statistics Institute of Sciences (APSIS), The George Washington University, Washington, DC 20052, USA}
\author{Matthew~G.~Baring}
\affiliation{Department of Physics and Astronomy, Rice University, MS-108, P.O. Box 1892, Houston, TX 77251, USA}
\author{Chryssa~Kouveliotou}
\affiliation{Department of Physics, The George Washington University, Washington, DC 20052, USA, gyounes@gwu.edu}
\affiliation{Astronomy, Physics and Statistics Institute of Sciences (APSIS), The George Washington University, Washington, DC 20052, USA}
\author{Zorawar~Wadiasingh}
\affiliation{Astrophysics Science Division, NASA Goddard Space Flight
  Center, Greenbelt, MD 20771}
\author{Daniela~Huppenkothen}
\affiliation{DIRAC Institute, Department of Astronomy, University of Washington, 3910 15th Avenue NE, Seattle, WA 98195, USA}
\author{Alice~K.~Harding}
\affiliation{Astrophysics Science Division, NASA Goddard Space Flight Center, Greenbelt, MD 20771}

\begin{abstract}

During a pointed 2018 \nustar\ observation, we detected a
  flare with a 2.2 hour duration from the magnetar
  1RXS~J170849.0$-$400910. The flare, which rose in $\sim25$ seconds
  to a maximum flux 6 times larger than the persistent emission, is
  highly pulsed with an rms pulsed fraction of $53\%$. The pulse
  profile shape consists of two peaks separated by half a rotational
  cycle, with a peak flux ratio of $\sim$2. The flare spectrum is
  thermal with an average temperature of 2.1~keV. Phase resolved
  spectroscopy show that the two peaks possess the same temperature,
  but differ in size. These observational results along with simple
  light curve modeling indicate that two identical antipodal spots,
  likely the magnetic poles, are heated simultaneously at the onset of
  the flare and for its full duration. Hence, the origin of the flare
  has to be connected to the global dipolar structure of the
  magnetar. This might best be achieved externally, via twists to
  closed magnetospheric dipolar field lines seeding bombardment of
  polar footpoint locales with energetic pairs. Approximately 1.86
  hours following the onset of the flare, a short burst with its own
  3-minute thermal tail occurred. The burst tail is also pulsating at
  the spin period of the source and phase-aligned with the flare
  profile, implying an intimate connection between the two
  phenomena. The burst may have been caused by a  magnetic
  reconnection event in the same twisted dipolar field lines anchored
  to the surface hot spots, with subsequent return currents supplying
  extra heat to these polar caps.

\vspace{25pt}
\end{abstract}

\section{Introduction}
\label{Intro}

Magnetars are a special subset of the isolated neutron star
family. Most exhibit long spin-periods (2-12~s) and large spin down
rates ($10^{-13}$-$10^{-11}$~Hz~s$^{-1}$), implying dipole magnetic
field strengths of the order of $10^{14}$~G
\citep{kouveliotou98Nat:1806}. They are persistent X-ray emitters,
with quasi-thermal spectra in the soft X-ray band and unique
non-thermal hard X-ray tails extending beyond 100~keV. Given that the
rotational energy loss rates of magnetars are typically $\sim
10^{33}$-$10^{34}$~erg s$^{-1}$, and therefore well below their soft
and hard X-ray luminosities ($\gtrsim 10^{35}$~erg s$^{-1}$), the high
energy emission of magnetars must be powered by a source other than
rotation. This source is widely believed to be the decay of their
large internal and perhaps external magnetic fields \citep[see][for
recent reviews]{olausen14ApJS:magCat,turolla15:mag,kaspi17:magnetars}. 

The most distinctive of the magnetar properties is their erratic
bursting activity, the most common of which are the short
($\sim0.2$~s), hard X-ray/soft $\gamma$-ray bursts with quasi-thermal
spectra. The peak luminosity of these bursts is in the range of
$10^{37}$-$10^{42}$~erg~s$^{-1}$, briefly dwarfing the persistent hard
X-ray signals \citep[e.g.,][]{collazzi15ApJS}. These short bursts are
at times followed by softer tails, lasting upward of few tens of
minutes \citep[e.g.,][]{gogus11ApJ:1806,an14ApJ:bursts}. In rare
occasions, short burst tails observed with RXTE exhibited a stronger 
persistent pulsed emission \citep[e.g.,][]{gavriil06ApJ:1048}. More
peculiar still, are the very energetic bursts observed from a handful
of magnetars, the intermediate and giant flares with peak energies of
the order of $10^{43}$~erg \cite[e.g.,][]{kouveliotou01ApJ:1900} and
upward of $10^{46}$~erg \citep[e.g.,][]{hurley99Natur}, respectively.
These events are usually followed by a decaying thermal tail of
duration a few hundred seconds pulsating at the pulse period of the
source. Finally, often accompanying these bursts, magnetars enter an
outburst episode, during which their underlying persistent emission is
spectrally and temporally altered, with brighter and harder X-ray
spectra, pulse profile changes, timing noise and/or glitches
\citep[e.g.,][]{archibald15ApJ:1048,camero14MNRAS,younes17ApJ:1806,
  younes17:1935}. The bursts may have an internal origin through
crust-quakes \citep{thompson95MNRAS:GF} or an external one caused by
magnetic reconnection, akin to solar flares \citep{lyutikov15MNRAS}.

1RXS~J170849.0$-$400910 (hereafter \src), is a magnetar with an 11
second period discovered with \asca\ \citep{Sugizaki97PASJ}.
Subsequent measurement of its spin down
$\dot{\nu}=-1.6\times10^{-13}$~Hz~s$^{-1}$ \citep{israel99ApJ} implied
a spin-down age of about 9~kyr and a surface polar magnetic field
strength of $B\approx 9.3\times10^{14}$~G. \src\ is one of the
brightest magnetars in the soft X-ray band with an absorbed 1-10~keV
flux $\sim4\times10^{-11}$~erg~s$^{-1}$~cm$^{-2}$. To date, it has
remained one of the few magnetars to not exhibit any outburst
activity in archival data \citep{olausen14ApJS:magCat}, except
  for a brighter hard X-ray pulsed flux during one archival \rxte\
  observation \citep{2014ApJ...784...37D}.

In this letter, we report on the first flaring activity from \src\
detected during a 2018 \nustar\ observation. Section~\ref{obs}
summarizes our observation and data reduction. We report here in
Section~\ref{res} only on the magnetar flaring results, for which we
will use the persistent signal from the source for comparison purposes
only; details of the persistent emission characteristics are deferred
to a later paper. The findings are summarized and discussed in 
Section~\ref{discuss}, focusing on the likely identification of two
hot polar caps as the sites for the flaring activity. Throughout the
paper we assume a distance to the source of 3.8~kpc
\citep{durant06ApJ}.

\section{Observations and data reduction}
\label{obs}

The {\it Nuclear Spectroscopic Telescope Array} (\nustar,
\citealt{harrison13ApJ:NuSTAR}) consists of two identical modules FPMA
and FPMB operating in the energy range 3-79~keV. \nustar\ observed
\src\ on 2018 August 28 for a total of 100.8~ks. We perform the data
reduction and analysis using \texttt{nustardas} version v1.8.0,
HEASOFT version 6.25, and the calibration files version 20190627. We
use the flags \texttt{saacalc=1, saamode=strict, tentacle=yes} to
correct for enhanced background activity visible at the edges of some
of the good time intervals immediately before or after entering the
South Atlantic Anomaly (SAA). This resulted in a total livetime
exposure of 92.8~ks. We extract source high-level science products,
using a circular region with a 60\arcsec\ radius, centered on the
brightest pixel around the source sky location. We use a circular
region of the same size on the same ccd as the source to extract
background light curves and spectra. We use the task
\texttt{nuproducts} to extract light curves and spectral files,
including ancillary and response files. We correct all light curves
for livetime, point spread function (PSF), and vignetting using
\texttt{nulccorr}. We also correct all arrival times to the solar
barycenter and to drifts in the \nustar\ clock caused by temperature 
variations \citep{harrison13ApJ:NuSTAR}.

We perform the spectral analysis using XSPEC version 12.10.1f
\citep{arnaud96conf}. We use the abundances of \citet{wilms00ApJ}, the
photo-electric cross-sections of \citet{verner96ApJ:crossSect}, and
the Tuebingen-Boulder interstellar medium absorption model
(\texttt{tbabs}) to account for X-ray absorption in the direction of
\src. We bin the spectra to have one count per bin and use the Cash
statistic (C-stat) in XSPEC for model parameter estimation and error
calculation, unless otherwise noted. We add a constant normalization to
all our spectral models to take into account any calibration
uncertainties between the two \nustar\ instruments, which we find to
be around $7\%$. This is larger than the typical $2\%$
cross-calibration accuracy, and may simply caused by the counting
statistics fully dominating the instrumental effects. Finally, we fix
the hydrogen column density in all spectra to $N_{\rm
  H}=1.89\times10^{22}$~cm$^{-2}$; the best fit value derived from the
persistent emission spectral analysis of our observation (Younes et
al. 2019 in prep.).

\begin{figure}[]
\begin{center}
\hspace*{-0.35cm}
\includegraphics[angle=0,width=0.49\textwidth]{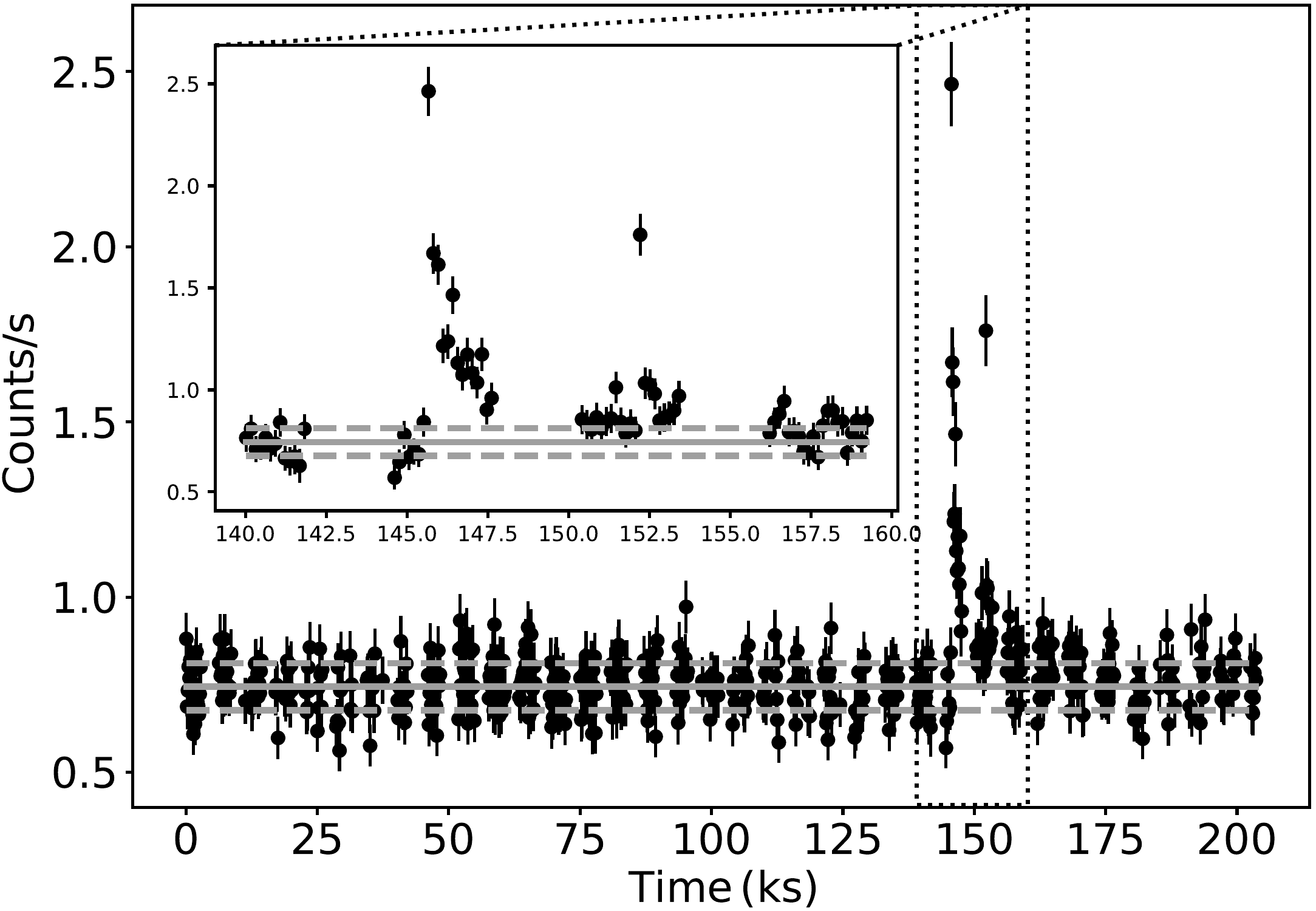}
\caption{\nustar\ FPMA+FPMB 3-20~keV light curve of \src\ binned at
  150~s resolution. The grey horizontal solid and dashed lines
  delineate the average of the persistent emission and its $1\sigma$
  uncertainty, respectively. The inset is a zoom-in of the dotted box
  region when enhanced emission above the persistent level is
  observed. See text for more details.}
\label{lc150}
\end{center}
\end{figure}

\section{Results}
\label{res}

The \nustar\ FPMA+FPMB 3-20~keV lightcurve of the source binned at a
resolution of 150~s is shown in Figure~\ref{lc150}. Enhanced emission
is clearly present at around 147~ks from the start of the observation.
The grey solid and dashed lines represent the average of the
persistent emission and its $1\sigma$ uncertainty as derived from the
beginning of the observation until the last Good Time Interval (GTI)
before the start of the enhanced emission. The inset is a zoom-in of
the dashed box-region around the time of the excess emission. The
\nustar\ background level in the same energy range is at the 1.5\%
level of the persistent emission. Statistical inspection of the enhanced
emission using a signal-to-noise method \citep[e.g.,][]{
  kaneko10ApJ:1550}, at varying time resolutions, reveals a long 
flare-like event as the onset of the activity, and a short burst with
its own tail 1.86~hours later. In the following sections, we first
present the detailed analysis of the earlier enhanced emission, while
we present the analysis of the short burst and its tail thereafter.

\subsection{Early enhanced emission - Flare}

\subsubsection{Timing analysis}
\label{flaretimAna}

The earlier enhanced emission was visually inspected at multiple
time-resolutions, starting at 0.128~s and up to 65~s. While we do find
strong variability starting at 0.5~s time-scales, we find no evidence
of an impulsive, short-like burst at the start of this enhanced
emission, and the peak is only resolved at the $\gtrsim8$~s
resolution. Hence, we refrain from labelling this event as a burst to
differentiate it from the typical magnetar short bursts; for the
purposes of this paper we instead refer to it as a flare.

\begin{figure}[]
\begin{center}
\hspace*{-0.35cm}    
\includegraphics[angle=0,width=0.49\textwidth]{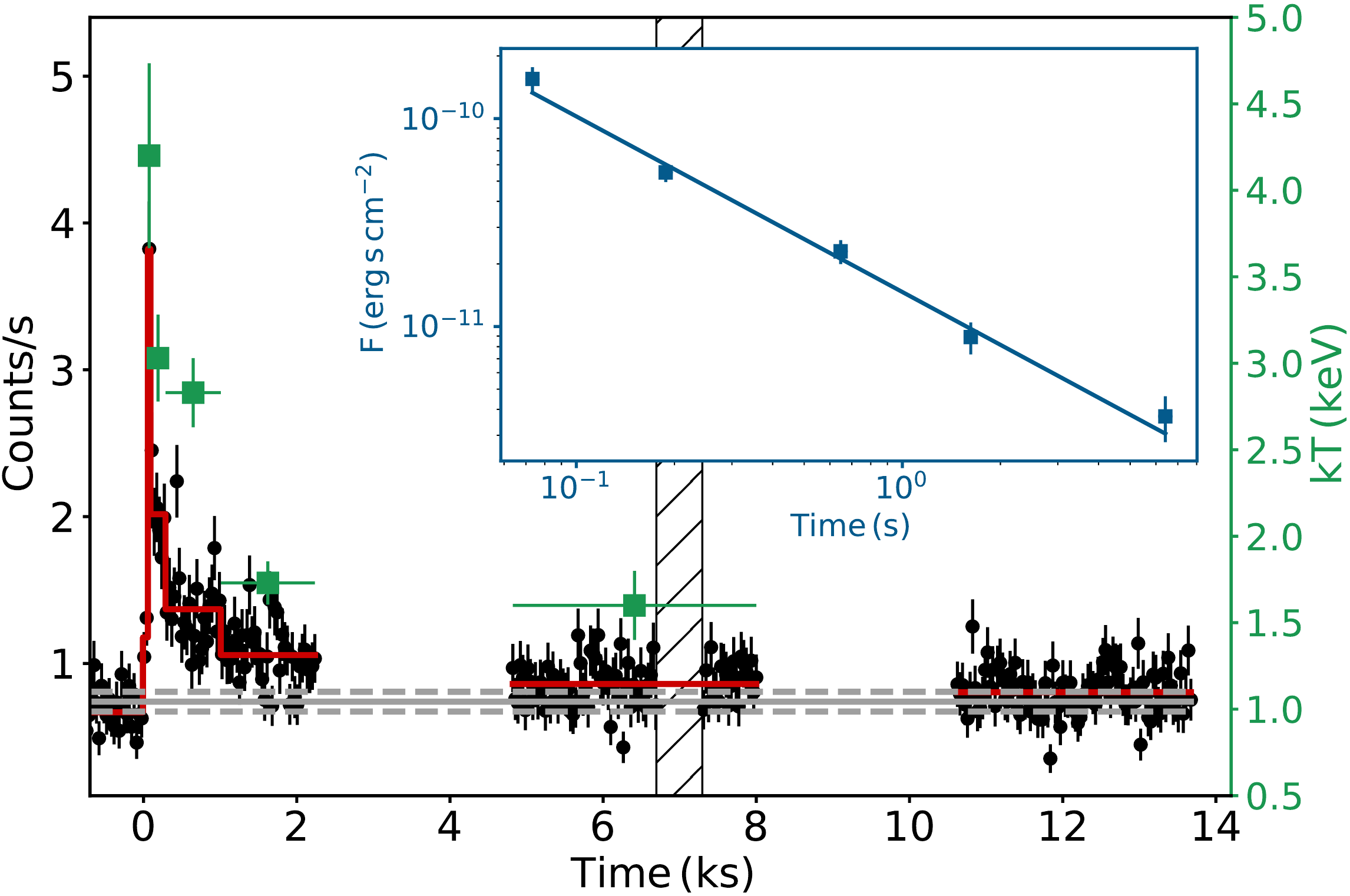}
\caption{Flare FPMA+FPMB 3-20~keV light curve shown
  at the 32~s resolution. The grey horizontal lines (persistent emission) are as in
  Figure~\ref{lc150}. The hatched region is a 400~s interval encapsulating the burst and its tail and was excluded from the flare analysis. The red stair curve is the Bayesian blocks
  representation of the flare. The right green axis and the green
  squares represent the BB temperature from time-resolved spectroscopy
  of the flare. The inset is the corresponding 3-20~keV
  absorption-corrected flux. The line is the best fit $t^{-\alpha}$ 
  PL decay of the flux with an index $\alpha=0.84\pm0.06$.}
\label{flareLC}
\end{center}
\end{figure}

Figure~\ref{flareLC} shows the 3-20~keV light curve encapsulating 3
long GTIs, the first of which includes the start of this enhanced
emission (we note that the flare is not detected beyond 20~keV). The
grey lines denote the persistent emission average and its 1$\sigma$
uncertainty. We exclude a 400~s interval during the second GTI, i.e.,
the hatched region, that covers the short burst and its tail. The red
solid stair curve is the Bayesian blocks representation of the flare
\citep{scargle2013apj:BB}. Emission above the level of the persistent
one is seen up until the end of the 2nd GTI, 8~ks after the start of
the flare. By the start of the third GTI, 2.6~ks later, the emission
seems to have already declined back to within $1\sigma$ of the
persistent emission.

To characterize the temporal properties of the flare, we fit the light
curve shown in Figure~\ref{flareLC} to 3 exponential functions, 1 for
the rise and 2 for the decay \citep[e.g.,][]{gavriil11ApJ:0142}. The
fit is good with a $\chi^2_\nu$ of 1.1 for 265 degrees of freedom
(dof). The properties are summarized in Table~\ref{timParam}. We find
a characteristic time-scale for the rise $\tau_{\rm r}=24\pm3$~s, an
initial decay $\tau_{\rm d}=52\pm7$~s, and a shallower one $\tau_{\rm
  s}=1180\pm180$~s. We find a duration $T90=3706\pm316$~s.

\begin{table}[t!]
\caption{Temporal parameters}
\label{timParam}
\vspace*{-0.5cm}
\begin{center}
\resizebox{0.49\textwidth}{!}{
\hspace*{-1.5cm}
\begin{tabular}{l c c c c c}
\hline
\hline
 \T\B & $T_0$ & $\tau_{r}$ & $\tau_{d}$ & $\tau_{s}$ & $T90$ \\
 \T\B & MJD & s & s & s & s\\
\hline
\hline
Flare \T\B &  58360.33620 & $24\pm3$  & $54\pm7$ & $1180\pm184$ & $3706\pm316$ \\
Burst \T\B & 58360.41281 & $0.06\pm0.01$ & $0.04$ & $0.58\pm0.03$ & $2.0$ \\
Tail \T\B &  58360.41287 & $3$ & $-$ & $43\pm5$ & $137$ \\
\hline
\hline
\end{tabular}}
\end{center}
\vspace*{-0.2cm}
{\bf Notes.} Parameters are estimated using a 32~s, 32~ms, and 16~s
resolutions for the flare, the burst, and the burst-tail respectively. 
Parameters without error bars imply a $68\%$ upper-limit. In such case
we do not quote any error on $T90$.
\end{table}

\begin{figure*}[]
\begin{center}
\includegraphics[angle=0,width=0.49\textwidth]{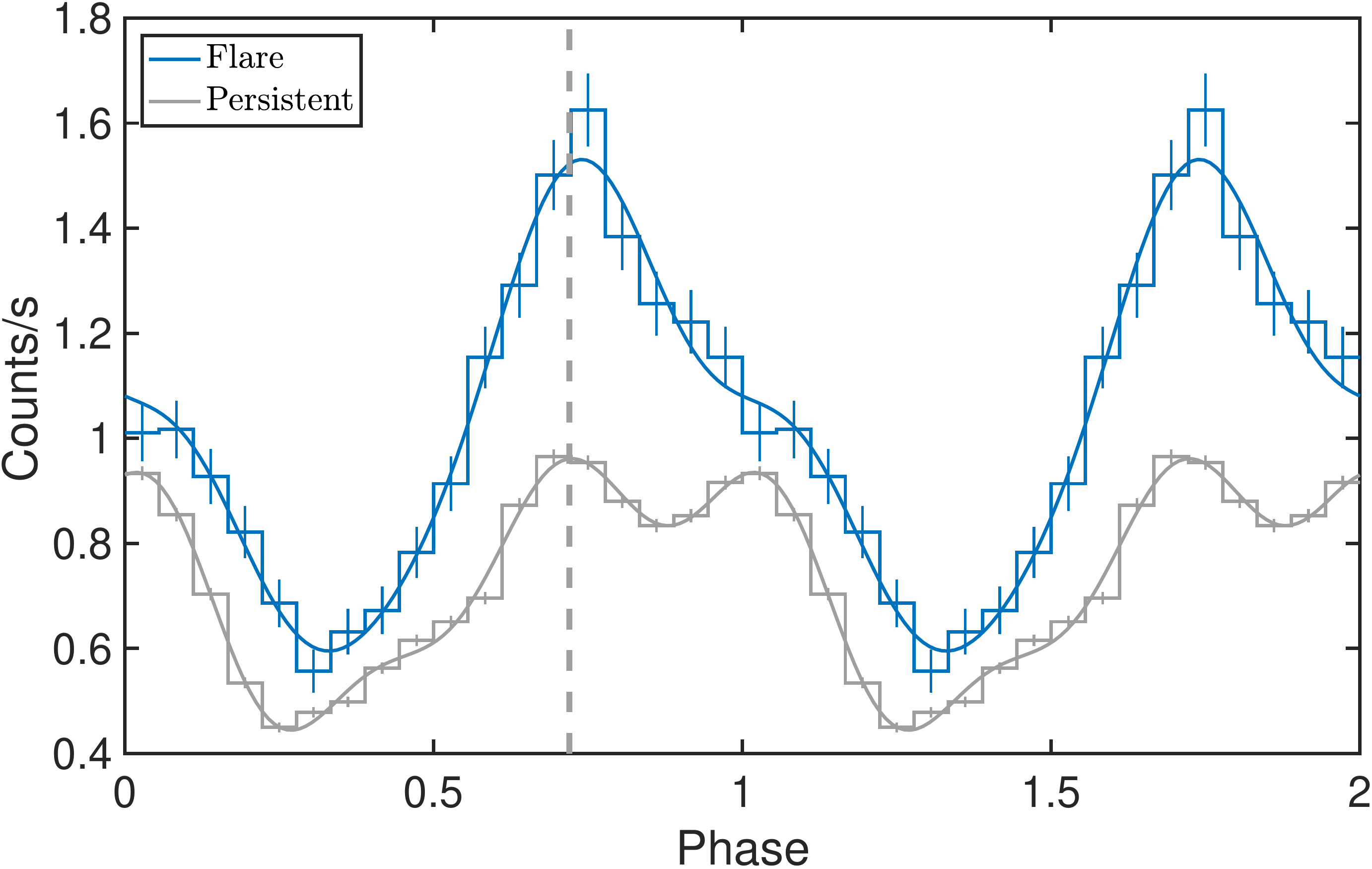}
\includegraphics[angle=0,width=0.49\textwidth]{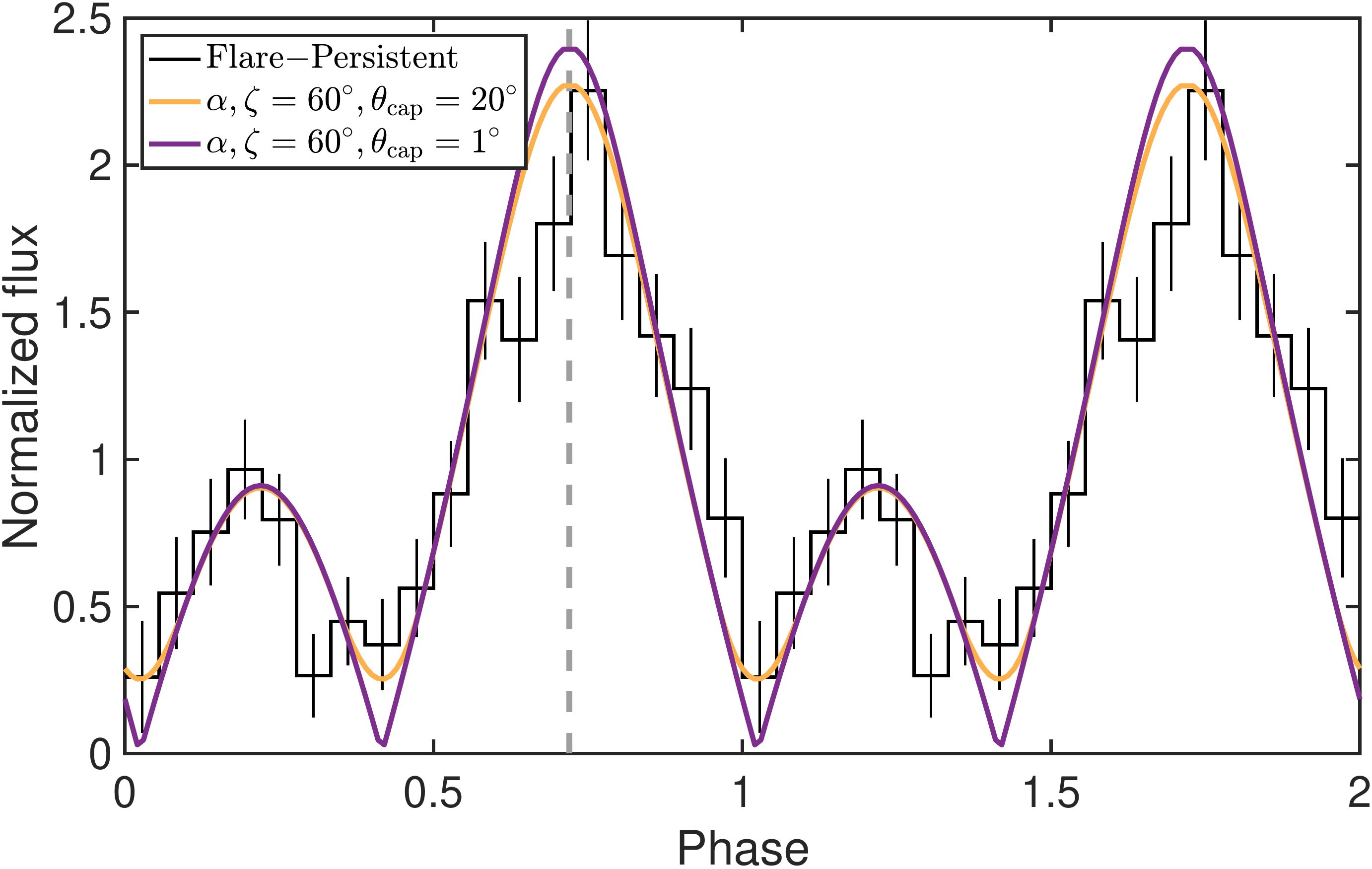}\\
\caption{{\sl Left panel.} Flare and persistent emission pulse
  profiles shown as blue and grey histograms, respectively, along with 
  $1$ sigma error bars. The solid
  lines are Fourier fits including the first three harmonics. {\sl
    Right panel.} Persistent-subtracted flare pulse profile, with the
  y-axis being counts normalized to the phase-averaged value.  The light orange
  and purple curves are the profiles from a simple two-pole hot spot model
  with cap sizes $\theta_{\rm cap}=20^{\circ}$ and $1^{\circ}$, respectively.  
  The model pulse profiles are for magnetic inclination $\alpha=60^{\circ}$. In both panels, the vertical dashed line is at $\phi_{\rm p}=0.72$, the phase of the first peak of the persistent emission pulse profile.}
\label{flarePP}
\end{center}
\end{figure*}

The flare pulse profile folded at the pulse frequency from the source
(0.09081755(3), Younes et al. in prep.) is shown in blue in
Figure~\ref{flarePP}. For comparison purposes, we also show the
persistent emission pulse profile in grey. The solid lines overlaid on
both curves are a Fourier series fit that includes contribution from
the first three harmonics. The difference in the pulse profile shapes
of the flare and the persistent emission implies that they are both
pulsed. Assuming that the persistent emission temporal (and spectral)
properties did not change during the flare emission\footnote{It would
  be quite drastic for the persistent emission to have changed during
  the flare period given that the pulse profiles of the GTIs just
  after and before the flare were consistent with one another, with
  the first 100~ks of the observation, and with the historical shape
  as measured through years of observations
  \citep{denhartog08AA:1rxs1708}.}, we subtracted the persistent
emission profile from the flare one (i.e., assuming that the
persistent emission is the background), which resulted in the profile
shown in the right panel of Figure~\ref{flarePP}. We derive an rms
pulsed fraction for the flare-only pulse profile of $53\pm5\%$.

The flare pulse profile consists of two peaks with different
brightness. We fit two Gaussians to the pulse profile which resulted
in a good fit with $\chi^2_\nu$ of 0.7 for 11 dof (note that a
  single Gaussian fit results in $\chi^2_\nu$ of 1.8 for 14 dof with
  strong residuals at phases $<0.5$). From this model, we derive a
peak separation of $0.55\pm0.04$~cycles, i.e., half a rotation.
  The main and subsidiary peaks are detected at rotational phases
  $0.74\pm0.02$ and $0.18\pm0.03$, respectively. The brighter peak is
aligned with the first peak of the persistent emission profile; the
peak that dominates the emission below few keV \citep{
  denhartog08AA:1rxs1708}, and is thermal in nature. We also searched
for any variability in the pulse profile with energy. We built pulse
profiles in two energy ranges, 3-8 and 8-20~keV. We find no
discernable difference in shape. The pulse fraction at high energies
is slightly larger at $60\pm9\%$ compared to $51\pm7\%$ at low
energies.

We also performed time-resolved pulse profile analysis. We folded the
light curve in each Bayesian block of Figure~\ref{flareLC}, starting
from the peak and up to 8~ks, at the source pulse period. The pulse
profile shape evolves from a broad pulse encompassing both peaks, to a
double peaked profile. Given the low statistics of each pulse profile,
measuring the pulsed fraction resulted in large error bars. Hence, we
merged the first three bins and the last two to construct two
relatively high signal-to-noise pulse profiles. We find that the
pulsed fraction at earlier times is $38\pm6\%$, markedly lower
compared with the last two Bayesian-block bins which returned a pulsed
fraction of $72\pm9\%$. Finally, we searched for Giant-flare like
transient quasi-periodic oscillations in the range 18 Hz to about 625
Hz \citep{israel2005,huppenkothen2014} during the flare. We find no
significant detection. That may be partially due to the very low
signal-to-noise data in the flare for such an analysis, which also
hindered any meaningful upper-limit measurement.

\subsubsection{Spectral analysis}
\label{flareSpeAna}

We fit the 4-20~keV time and phase averaged, persistent-emission
corrected, flare spectrum with a power-law (PL) and a blackbody (BB)
model. Given the relatively large total number of flare counts
($\sim1500$ FPMA+FPMB), we group the spectra to have a minimum of 70
counts per bin and used the $\chi^2$ statistic. The BB model resulted
in a better fit compared to the PL model with a $\chi^2_\nu$ of 1.03
and 1.3, respectively, for 44 dof. We find a BB temperature
$kT=2.14\pm0.12$~keV and a 3-20~keV flux $F_{\rm
3-20~keV}=(1.00\pm0.07)\times10^{-11}$~erg~s$^{-1}$~cm$^{-2}$, which 
is $36\%$ of the persistent emission flux in the same energy range.
Assuming circular geometry, we derive a phase-averaged radius for the
emitting region $R=84_{-9}^{+11}$~m. We summarize the best fit
parameters of both models in Table~\ref{specParam}. 

For spectroscopy including the higher energy window above 20 keV,
  we added a PL component to the BB one, and thereby derived an
  upper-limit on the detection of a putative hard X-ray tail component
  for the flare. We fixed the PL $\Gamma$ to 0.7, similar to the one
  we derive for the persistent emission during this observation. The
  BB temperature and normalization were left free to vary. We obtained
  an upper limit $F_{\rm
    20-70~keV}\lesssim8\times10^{-12}$~erg~s$^{-1}$~cm$^{-2}$. This is
  $52\%$ of the 20-70 keV persistent flux. Given that the increase in
  the 3-20 keV flux during the flare was at the $36\%$ level, one
  cannot exclude the possibility for such an increase/flare to have
  occurred in the hard PL as well.

A time-resolved spectral analysis of the flare was performed by fitting a
BB model to each of the Bayesian block bins, starting from the
peak. This revealed a strong cooling trend throughout the flare
with the temperature decreasing from $4.2\pm0.5$~keV to
$1.6\pm0.2$~keV. On the other hand, no significant change appeared in
the area of the BB emitting region. The 3-20~keV flux decreases from
$(1.55\pm0.20)\times10^{-10}$~erg~s$^{-1}$~cm$^{-2}$ to
$3.7_{-0.7}^{+0.6}\times10^{-12}$~erg~s$^{-1}$~cm$^{-2}$. The flux
decay is well fit with a PL function $F(t)\propto t^{-\alpha}$ with
$\alpha=0.84\pm0.06$. The temperature and flux decay trends
are shown in Figure~\ref{flareLC}, while all spectral parameters are
summarized in Table~\ref{specParam}. We derive a flare fluence
$\Phi=(7.0\pm0.6)\times10^{-8}$~erg~cm$^{-2}$ and a total emitted
energy $E=(1.2\pm0.1)\times10^{38}$~erg, which is about $50\%$ of
the energy emitted in the persistent emission in the same time-span.

Finally, we carried out phase-resolved spectroscopic analysis of the
source. We divided the pulse profile into three bins encapsulating the
two peaks and the interpulse emission (Table~\ref{specParam}). The BB
temperature is consistent within $1\sigma$ between the two peaks,
whereas, the area of the emitting region is a factor of 2.7 smaller in
the weaker pulse compared to the main one. The interpulse spectrum
exhibits a slightly smaller temperature than the pulses and an area
comparable to the size of the main pulse.

\begin{table*}[t!]
\caption{Spectral parameters}
\label{specParam}
\vspace*{-0.5cm}
\begin{center}
\resizebox{0.80\textwidth}{!}{
\hspace*{-1.5cm}
\begin{tabular}{l c c c c c c c c}
\hline
\hline
\T\B & Model & $kT$ & $\Gamma$ & $R$ & $F$ & $L$ & $\chi^2$/dof & Cstat/dof \\
\T\B &           & (keV) &                  & (m) & erg~s$^{-1}$~cm$^{-2}$ & erg~s$^{-1}$ & & \\
\hline
\hline
 \multicolumn{9}{c}{Flare}\\
\hline
0.0-1.0 \T\B & BB & $2.14\pm0.12$ & $-$ & $84_{-9}^{+11}$ & $10_{-0.8}^{+0.6}$ & $1.7\pm0.1$ & 45/44 & $-$\\
0.0-1.0 \T\B & PL & $-$ & $2.12\pm0.13$ & $-$ & $12\pm1$ & $2.1\pm0.2$ & 45/44 & $-$ \\
\hline
\multicolumn{9}{c}{Time-resolved spectroscopy}\\
\hline
Bin~1 \T\B & BB & $4.2_{-0.4}^{+0.6}$ & $-$ & $97_{-15}^{+23}$    & $155_{-22}^{+21}$ & $27\pm3$ & $-$ & 100/111 \\
Bin~2 \T\B & BB & $3.0_{-0.2}^{+0.3}$ & $-$ & $101_{-13}^{+17}$ & $55\pm6$ & $9.5\pm1.0$ &  $-$ & 235/268 \\
Bin~3 \T\B & BB & $2.8\pm0.2$        & $-$ & $76\pm10$        & $23\pm2$ & $4.0\pm0.3$ &  $-$ & 360/426 \\
Bin~4 \T\B & BB & $1.7_{-0.1}^{+0.2}$ & $-$ & $133_{-18}^{+27}$ & $9\pm1$ & $1.6\pm0.2$ &  $-$ & 407/430 \\
Bin~5 \T\B & BB & $1.6\pm0.2$        & $-$ & $100_{-20}^{+40}$ & $3.7\pm0.6$ & $0.64\pm0.10$ & $-$ & 585/593 \\
\hline
\multicolumn{9}{c}{Phase-resolved spectroscopy}\\
\hline
0.06-0.28 \T\B & BB & $2.3\pm0.2$  & $-$ & $64_{-11}^{+17}$ & $8_{-2}^{+1}$ & $1.4_{-0.3}^{+0.2}$ &  $-$ & 404/472 \\
0.56-0.94 \T\B & BB & $2.2\pm0.1$  & $-$ & $105\pm10$     & $17_{-1}^{+2}$ & $2.9_{-0.2}^{+0.4}$ &  $-$ & 472/575 \\
Rest \T\B          & BB & $1.8\pm0.2$  & $-$ & $98_{-19}^{+30}$ & $6.4_{-0.9}^{+0.8}$ & $1.1\pm0.1$ &  $-$ & 424/489 \\
\hline
\hline
\multicolumn{9}{c}{Burst}\\
\hline
0.0-1.0 \T\B & BB & $6.2_{-0.8}^{+1.2}$  & $-$ & $180_{35}^{+45}$ & $3200_{-600}^{+800}$ & $550_{-100}^{+140}$ &  $-$ & 60/61 \\
0.0-1.0 \T\B & PL & $-$ & $0.6\pm0.2$  & $-$ & $3900_{-800}^{+900}$ & $670_{-130}^{+160}$ &  $-$ & 54/61 \\
\hline
\multicolumn{9}{c}{Tail}\\
\hline
0.0-1.0 \T\B & BB & $2.9_{-0.4}^{+0.5}$  & $-$ & $80\pm20$ & $34_{-8}^{+7}$ & $6\pm1$ &  $-$ & 154/163 \\
0.0-1.0 \T\B & PL & $-$ & $1.7_{-0.2}^{+0.3}$  & $-$ & $53_{-0.7}^{+0.6}$ & $9\pm1$ &  $-$ & 155/163 \\
\hline
\end{tabular}}
\end{center}
\vspace*{-0.2cm}
{\bf Notes.} The first column represents the time bins
(Figure~\ref{flareLC}) or rotational phase intervals for which the
spectral analysis is performed. Fluxes are in units of
$10^{-12}$~erg~s$^{-1}$~cm$^{-2}$ and derived in the energy range
3-20~keV, except for the short burst, for which we report the 3-35~keV
flux. Radii and luminosities are derived by adopting a 3.8~kpc
distance \citep{durant06ApJ}, and the latter is in units of
$10^{34}$~erg~s$^{-1}$. All listed uncertainties are at the $1\sigma$
level.
\end{table*}

\subsection{Short burst and tail}
\label{burst2Ana}

\begin{figure*}[t!]
\begin{center}
\includegraphics[angle=0,width=0.51\textwidth]{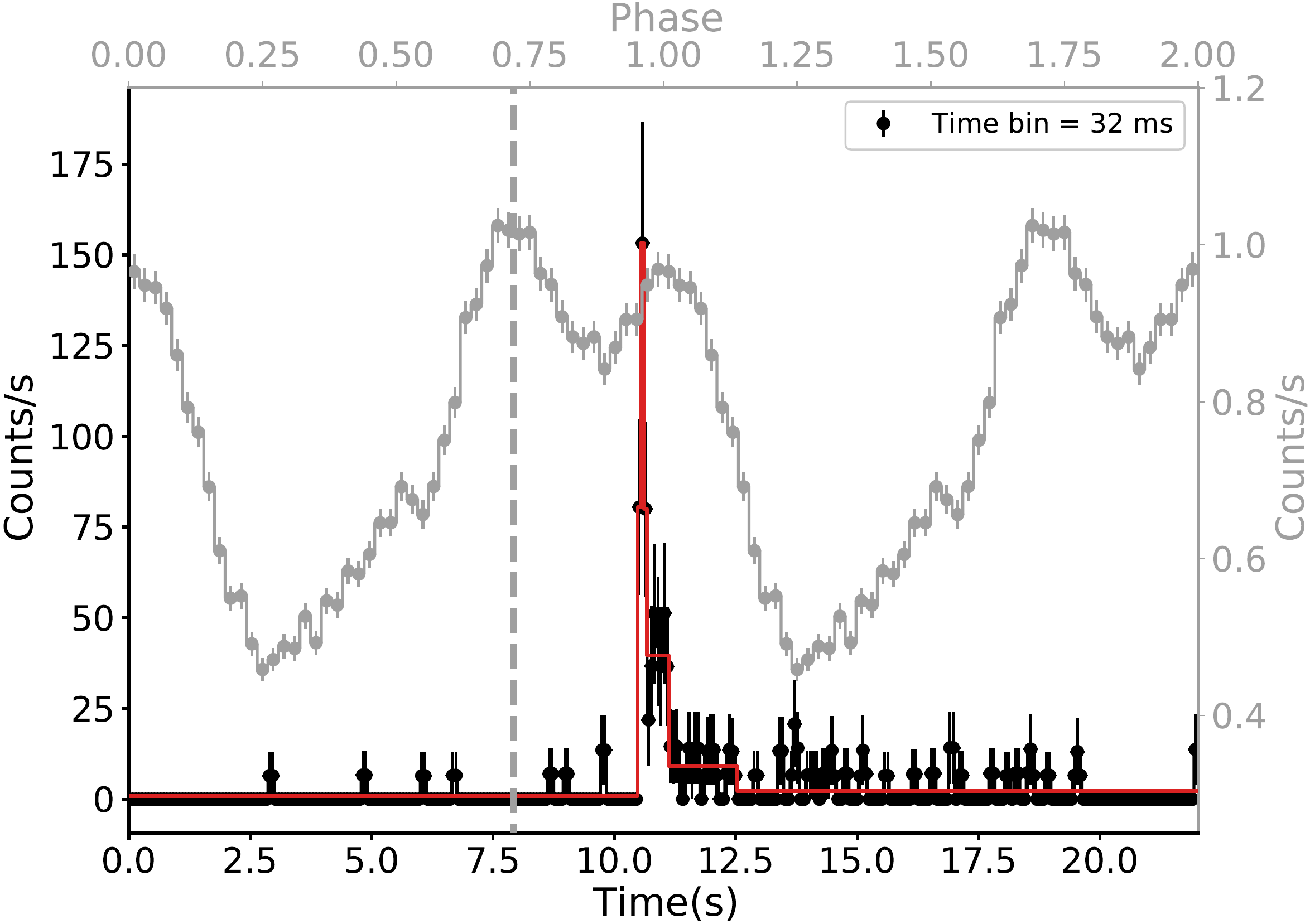}
\includegraphics[angle=0,width=0.48\textwidth]{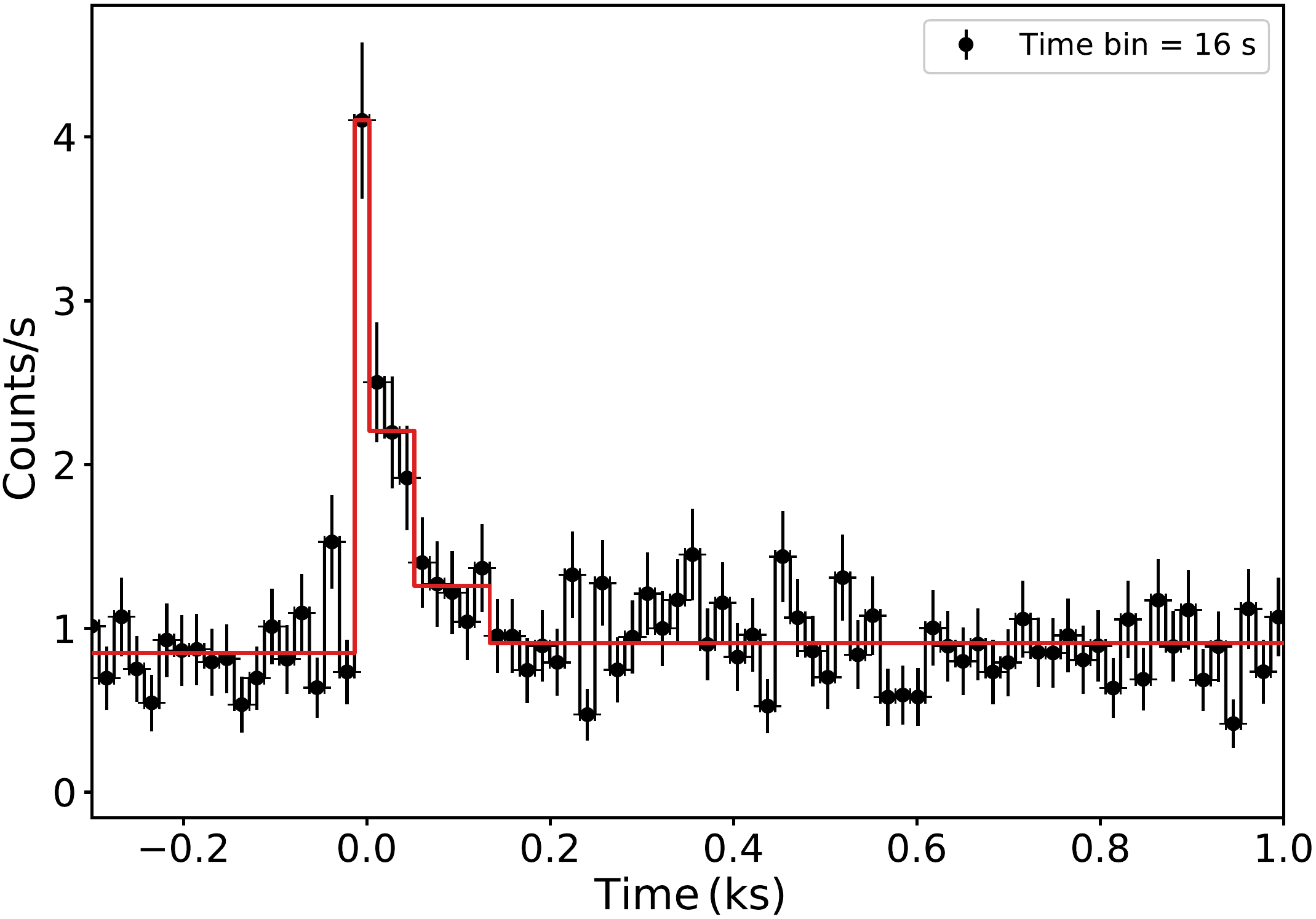}\\
\includegraphics[angle=0,width=0.465\textwidth]{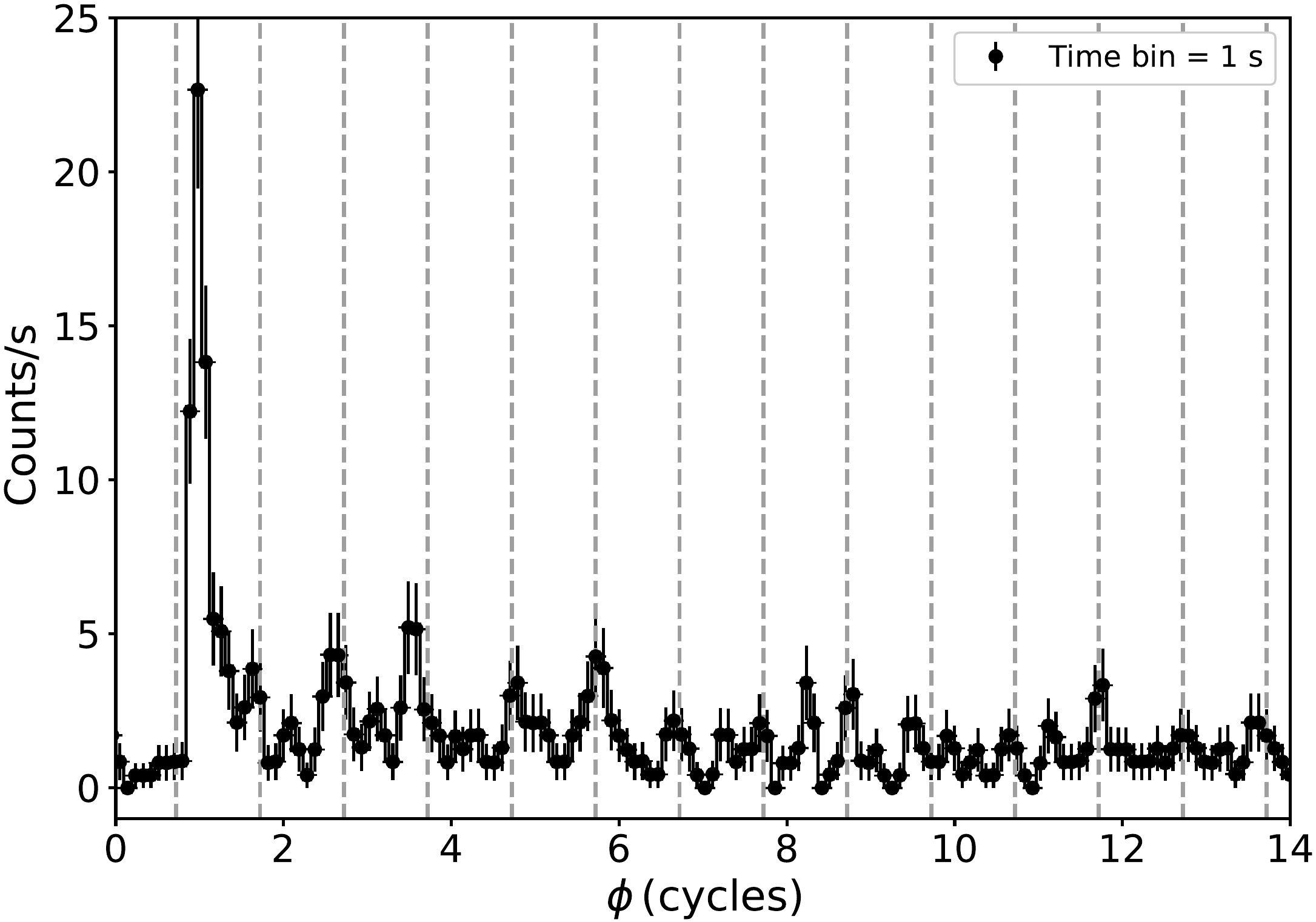}
\hspace{0.4cm}{
\includegraphics[angle=0,width=0.48\textwidth,height=2.28in]{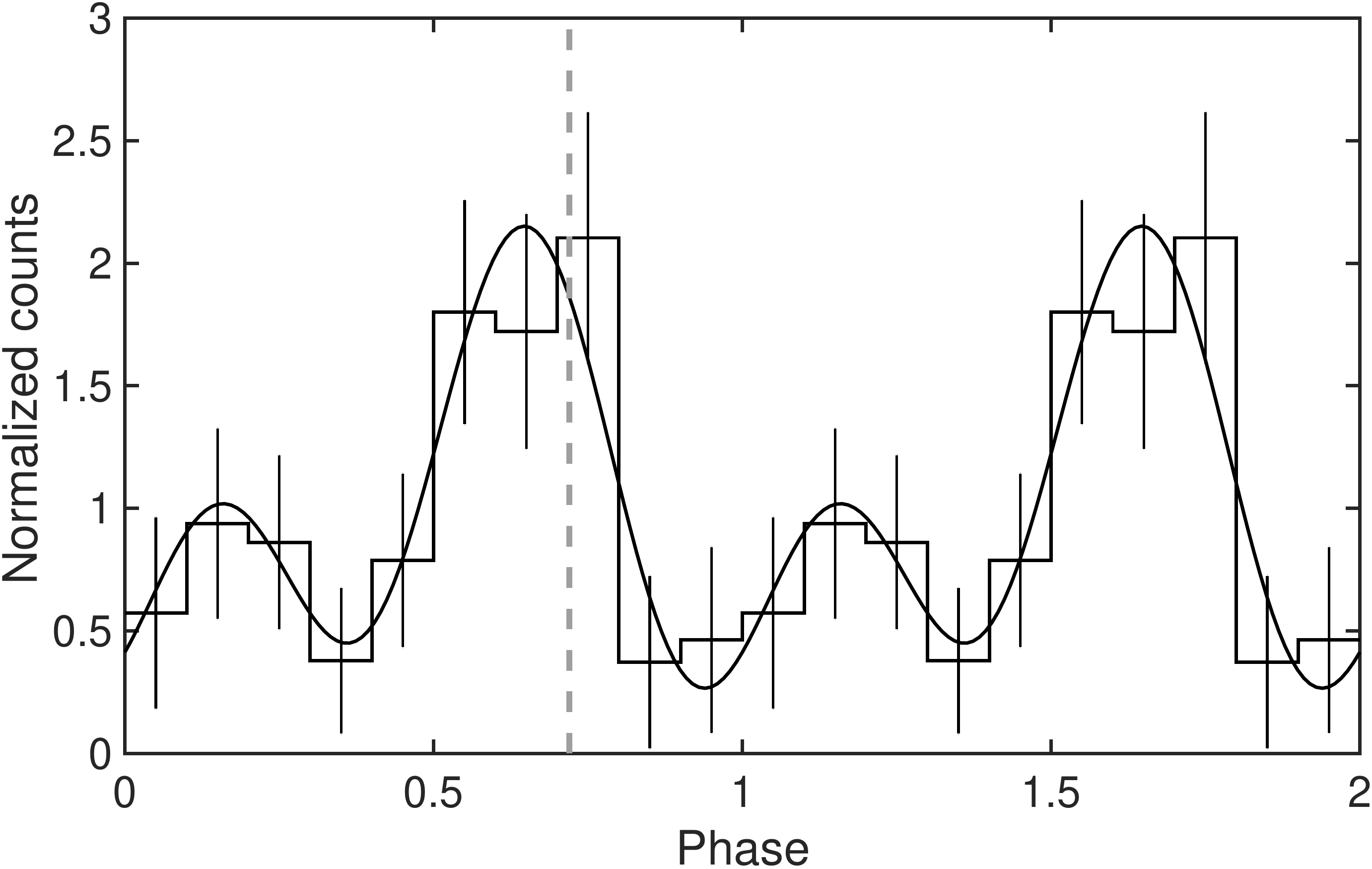}}
\caption{{\sl Upper-left panel.} Short burst light curve shown at the 32~ms time-scale. The red stair curve is the burst Bayesian blocks representation. The start time of the light curve corresponds to T0 at which phase=0, shifted by a time equivalent to 12915 rotations. The end time is 2 rotations later. The grey curve is a high resolution persistent emission pulse profile. Two rotations are shown. The grey vertical dashed line corresponds to $\phi_{\rm p}=0.72$. {\sl Upper-right panel.} Burst light curve shown at the 16~s time-scale, with its Bayesian blocks representation in red. {\sl Lower-left panel.} Burst light curve shown at the 1.0~s time-scale, 1/11th of \src\ rotational period. The x-axis is in rotational phase $\phi$ normalized by $2\pi$ (i.e., cycles). The start phase corresponds to $\phi$=0, shifted by 12915 rotations. The vertical dashed grey lines are drawn at $\phi=\phi_{\rm p}+n$, with $n$ an integer from 0 to 14. {\sl Lower-right panel.} Folded light curve 
 of the burst tail shown in the lower-left panel, excluding a 2~s interval around the burst time. Grey vertical dashed line corresponds to $\phi_{\rm p}$. See text for more details.}
\label{B2plots}
\end{center}
\end{figure*}

We plot the short burst light curve at the 32~ms time-scale in the
upper-left panel of Figure~\ref{B2plots}. The full time interval that
is shown spans two rotational periods. The start time of the light
curve corresponds to T0 at which phase=0 (see Section~\ref{res}),
shifted by a time equivalent to an integer number of rotations
(12915). The end time is two rotations later. To identify the phase at
which the burst occurred, we plot in grey two cycles of the persistent
emission pulse profile. The burst aligns with the second peak of that
profile, which fully dominates the emission at energies $>20$~keV. We
fit the burst light curve to 3~exponential functions using a
maximum-likelihood method. We measure $T90\approx2$~s. The right panel
of Figure~\ref{B2plots} shows the burst light curve at 16~s
resolution. Only 2 exponential functions are required to fit this
light curve, and we measure a $T90\approx140$~s. The temporal
properties of this burst and its tail are summarized in
Table~\ref{timParam}.

The light curve of the burst tail at 1~s resolution (one-eleventh of
the rotational period of \src) is shown in the lower-left panel of
Figure~\ref{B2plots}. Phase 0 corresponds to T0, shifted by 12915
rotations. The dashed vertical lines are at $\phi=\phi_{\rm p}+n$ with
$n$ an integer from 0 to 14. Peaks during the tail light curve are
clearly seen aligned at $\phi$, with a small hint of interpulses. We
folded the same light curve at the spin period from the source and
subtracted the persistent emission pulse profile. The result is shown
in the right panel of Figure~\ref{B2plots}. The pulse is detected at
the $4.8\sigma$ level. Fitting a two Gaussian model to the pulse 
profile, we find the phases of the main and secondary peaks at
$0.65\pm0.10$ and $0.18\pm0.10$, respectively. These phases
are within 1$\sigma$ uncertainty from the flare pulse peaks
(Section~\ref{flaretimAna}). Moreover, their peak flux ratio of
about 2 is also consistent with that of the flare. While the
secondary peak is not, on its own, statistically significant, its
appearance at the same rotational phase as the minor peak in the
flare pulse profile and at the same flux level enhance confidence
in the reality of its existence.

We fit the short-burst 3-40~keV phase-averaged spectrum with a BB and
a PL model. The former results in a CSTAT of 60 for 61 dof. We find a
BB temperature $kT=6.2_{-0.8}^{+1.2}$ and a radius for the emitting
region $R=180_{-35}+{45}$~m. The BB 3-40~keV flux is
$3.2_{_-0.6}^{+0.8}\times10^{-9}$~erg~s$^{-1}$~cm$^{-2}$.  The PL
model results in a slightly better fit with a CSTAT of 54 for 61
dof. We find a photon index $\Gamma=0.6\pm0.2$ and a 3-40~keV flux of
$3.9_{_-0.8}^{+0.9}\times10^{-9}$~erg~s$^{-1}$~cm$^{-2}$. This implies
a burst fluence of about $7.8\times10^{-9}$~erg~cm$^{-2}$. Bursts
analyzed with \swift/BAT or \fermi/GBM usually require either a cutoff
at high energies or, more accurately, are best fit with a 2 BB model
\citep{lin12ApJ:1550,younes14ApJ:1550}. This extra component is not
required in our data, which could be attributed to the lower
sensitivity of \nustar\ at energies beyond 30~keV. Finally, we fit the
4-20~keV tail spectrum with a PL and a BB model. The PL model results
in CSTAT of 155 for 163 dof. We find a photon index $\Gamma=1.7\pm0.2$
and a 3-20~keV flux of
$3.5_{-0.8}^{+0.6}\times10^{-11}$~erg~cm$^{-2}$~s$^{-1}$. The BB model
gives a CSTAT of 154 for 163 dof, a phase-averaged temperature
$kT=2.9\pm0.4$~keV with radius $R=80\pm20$~m. The former is slightly
larger than the temperature in the flare, whereas both possess a
similar area. We find a 3-20~keV flux
$F=(2.9\pm0.6)\times10^{-11}$~erg~cm$^{-2}$~s$^{-1}$. The fluence in
the tail of the burst is $\Phi_{\rm t}=5.2\times10^{-9}$~erg~cm$^{-2}$.
Hence, the energy emitted by \src\ during the burst and its tail is
$\sim1.8\times10^{37}$~erg; an order of magnitude smaller than the
total energy emitted during the flare.

\section{Summary and discussion}
\label{discuss}

In this Letter, we report on flaring activity from the magnetar \src,
which, until now, had not shown any of the bursting behaviors common
to this class of sources. The transient activity lasted about 2.2
hours with $T_{90}\approx1$~hr and a rise time of $\tau_{\rm r}=24$~s.
It was not triggered by any impulsive event, such as a short burst,
and no unusual activity was apparent prior to the flare. Such enhanced
emission in magnetars has been seen after giant and intermediate
flares \citep[e.g.,]{hurley99Natur,ibrahim01ApJ:1900,palmer05Natur},
short milli-second bursts \citep{woods05ApJ:xte1810,gavriil11ApJ:0142,
  an14ApJ:bursts}, and at the onset of a major bursting episode
\citep{kaneko10ApJ:1550}. Hence, the enhanced emission discussed here
demonstrates that long-duration, low-level activity could also happen
in isolation, and it may be quasi-continuous in at least some
magnetars \citep[see also][]{esposito19AA:rcw103}.

A unique aspect to the \src\ flare is its highly pulsed nature,
  with a pulse profile differing in shape from the persistent emission
  one. It consists of two peaks separated by half a cycle, with a flux
  ratio of two. The brighter peak coincides in phase with the
  quasi-thermal, soft emission ($\lesssim4$~keV) peak of the
  persistent pulse profile \citep[e.g.,][]{denhartog08AA:1rxs1708}.
  Its rms pulse fraction is $53\%$; a factor two larger than the
  persistent emission one ($PF=28\%$). Finally, the two peaks in the
  flare pulse profile are present throughout the whole episode of
  activity, and with similar flux ratio.

The time-integrated, phase-averaged flare spectrum is best fit with a
thermal blackbody model of temperature $kT=2.1$~keV, which is 4.5
times higher than the persistent emission blackbody temperature
\citep[e.g.,][]{rea05MNRAS:1708}. The effective area of the emitting
region, with a radius $R\approx100$~m, is much smaller than that of the
persistent emission \citep[$R_{\rm
  per}\approx2$~km,][]{rea05MNRAS:1708}. We cannot exclude the
  presence of a similar transient hard X-ray component in our data if
  the increase is at a similar level as the 3-20~keV flux.
  Phase-resolved spectroscopy of the flare reveals that both peaks are
  thermal with similar BB temperatures of about 2.1~keV. They only
  differ in the apparent size of their emitting area.

The flare double-peaked pulse profile with peak separation of half a
rotation (Fig.~\ref{flarePP}), the appearance of the two peaks at the
flare onset and for its full duration, and the peaks similar BB
temperature strongly suggest simultaneous heating to two antipodal
surface spots with the same population of energetic particles.The
most natural regions on the magnetar surface are the two magnetic
polar caps, since areas connected to higher-order fields are usually
threading footprints with closer proximity on the surface. To test this hypothesis,
we developed a simple two-polar hot spot model for generating pulsed
emission profiles.  The {\sl identical} surface spots were centered at
opposite poles, and spanned magnetic colatitudes between zero and
$\theta_{\rm cap}$, and $\pi$ and $\pi - \theta_{\rm cap}$.  Results
for $\theta_{\rm cap}=1^{\circ}, 20^{\circ}$ are illustrated in
Fig.~\ref{flarePP}, though profiles for a full range of  $0 <
\theta_{\rm cap} \leq \pi /2$ were surveyed.

The hot spots emitted uniformly across their surfaces.  
A simple anisotropic and azimuthally-symmetric
emission flux profile of ${\cal F}(\theta_z) = {\cal F}_0 (1 +
\sin^2\theta_z)^{-\beta} $ was assumed at each point on each cap, with
$\theta_z$ representing the local zenith angle.  This choice is
motivated by general expectations of radiative transfer in
highly-magnetic neutron star atmospheres. The zenith flux ${\cal F}_0$
was fixed for all pulse profile simulations, and the anisotropy index
$\beta$ allowed to vary from $\beta =0$ (isotropy) to $\beta = 4$
(pencil-beam like).  The two other key parameters for controlling
pulse profile morphology are the angle $\alpha$ between the magnetic
dipole and rotation axes, and the angle $\zeta$ of the observer
viewing direction to the magnetar rotation axis.  The simulations were
performed for flat spacetime to aid expediency of the analysis.

A broad array of pulse profiles was thus generated.
The special cases of an orthogonal rotator ($\alpha = 90^{\circ}$) or
equatorial viewing ($\zeta = 90^{\circ}$) generate two identical peaks
that do not match the observations.  Low values of $\zeta \lesssim
50^{\circ}$ occult the antipodal (remote) polar cap so that only a
single peak emerges.  The relative heights of the two peaks (each
symmetric) and their widths were used to constrain the
parameters $\alpha$, $\zeta$ and $\beta$.  This survey revealed that
$\alpha\sim 60^{\circ}$ and $\zeta \sim 60^{\circ}$ as displayed in
the right panel of Fig.~\ref{flarePP} generated the best fit, with a
reduced $\chi^2$ of $1.7$ for $\theta_{\rm cap} = 20^{\circ}$ (orange curve),
and $\chi^2$ of 2.2 for $\theta_{\rm cap} = 1^{\circ}$ (purple).  
The tolerance in these values was around $\pm 10^{\circ}$, albeit with an
anti-correlation between $\alpha$ and $\zeta$ choices.  The best-fit
$\beta$ index was $1/2$, indicating very modest surface anisotropy,
though we note that the larger values of $\beta = 1-2$ narrowed the
peaks only modestly.  It is evident that reducing the $\theta_{\rm cap}$ value 
to around $1^{\circ}$ narrows the peaks slightly and increases the pulse
fraction somewhat.  Note that the $\theta_{\rm cap}=1^{\circ}$ case lowers 
the effective spot radius to $\sim 174$~m (for a stellar radius of 10 km),
i.e., much closer to that estimated from the observed flux.  

The quality of the fit is sufficient to indicate approximate
consistency of the two-polar cap scenario with the observed pulse
profile.  Refined treatment of curved photon trajectories is expected
to modify the fit parameters only by fairly small fractions. The main
general relativistic manifestations are to increase the effective
brightness of the antipodal cap somewhat and enlarge the range of
phases for which this cap is visible. These influences enhance and
broaden the subsidiary peak slightly so as to effectively reduce the
pulsed fraction. A more detailed modeling of the flare light curve
including all the above is deferred to future work.

The large value of $\alpha$ obtained here contrasts that found for 
1E 1841-045 
\citep[$\alpha \approx 15^{\circ}$][]{An2015ApJ} and also $\alpha \sim
0^{\circ}$ identified for \src\ \citep{Hascoet2014ApJ} from 
{\it spectral} modeling of {\it magnetospheric hard X-ray tail emission}.  
Results in those two papers emphasize phase-averaged spectral fitting 
using the magnetic Thomson scattering model of \cite{beloborodov13ApJ}.  
The strong dependence of such resonant upscattering spectra on pulse phase
\citep[e.g.,][see also Fig.~7 of Beloborodov 2013]{wadiasingh18ApJ} 
renders spectral similitude a relatively 
poor probe of the geometric angles $\alpha$ and $\zeta$. Pulse profile
comparison between data and model is a more precise diagnostic for 
these angles.  Such was emphasized in the flat spacetime pulse 
profile fitting of XMM data below 10 keV performed by \cite{Albano2010ApJ} 
for two magnetars.  They found that $\alpha\sim 20^{\circ}$ for 
XTE J1810-197 with its simple, single-peaked pulsation, and
$\alpha\sim 85^{\circ}$ for CXOU J164710.2$-$455216 with its
multi-peaked pulse profile; more along the lines of the geometry
constraints derived in this paper.

The main question now is, how is it possible to heat both magnetic
polar caps simultaneously at the onset of the flare and for the length
of its decay? While this can potentially happen in subsurface zones, 
the observed antipodal heating demands that the site for activation 
has a prompt/simultaneous physical connection to both poles. This is
likely difficult to achieve below the surface given the diffusive nature 
of heat transport.  Collective energy transfer mediated by acoustic or
Alfv\'enic modes may provide an alternative possibility, but would
require a specialized transport geometry below the surface.

For a magnetospheric origin, dipolar fields may undergo large twists, 
perhaps due to a shift in the underlying crustal area to which they are
anchored. The toroidal components to these fields generate strong electric
fields that rapidly accelerate magnetospheric electrons ejected from the
surface, precipitating prolific pair creation. The resultant
over-twisting is unsustainable for long, and will naturally relax with
an evolution progressing towards quasi-polar field-line footpoints
\citep{Chen+Belo2017ApJ}. Pairs accelerated in the twisted field zones
bombard the polar surface regions \citep[e.g.,][]{beloborodov16:heat,
  gonzalez19MNRAS}, so that accompanying enhanced heating of both
polar caps simultaneously is naturally expected. Twists in dipolar
magnetic fields, however, are also believed to power non-thermal hard
X-ray emission \citep[e.g.,][]{beloborodov13ApJ,wadiasingh18ApJ}.
Assuming that the increase in this component is at the level of the
soft thermal emission, its non-detection is then expected given the
upper-limit we derive on the 20-70~keV flare flux
(Section~\ref{flareSpeAna}).

The temporal and spectral properties of the short burst detected from
\src\ towards the end of the flare is typical of magnetar short bursts
observed with \nustar\ \citep[e.g.,][]{an14ApJ:bursts}. The burst time
is phase-locked with the second peak of the persistent pulsed emission which
completely dominates above $20$~keV \citep[Figure~\ref{B2plots},][]{
  denhartog08AA:1rxs1708}. It occurs at the flare pulse profile phase
minimum, and so may not be associated with either pole. The short
burst is followed by a 3~minute tail, well fit with a BB model having
phase-averaged $kT\sim 3$~keV and $R \sim 80$~m. The temperature is
larger than the one derived for the flare spectrum, while the radii
are consistent. Moreover, the tail is pulsating at the spin period of
the source, with a pulse profile aligning in phase with that of the
flare and exhibiting a similar pulsed fraction.

The above common characteristics between the flare and the burst tail point 
to an intimate connection between the origin of both events. They are
naturally explained if the burst also occurred in  the magnetosphere,
on the same field loops anchored to the initial heated magnetic polar
cap areas. These loops experienced a twist at the onset of the
activity that perhaps precipitated a magnetic reconnection event
\citep{lyutikov15MNRAS}, causing the burst. Return currents deposit
extra heat onto the same surface region, resulting in an increase in
the surface  temperature without impacting the size of the emitting
area. In summary, the flare observations persent here provide strong
evidence for a twist to the global dipolar magnetic field as the
source of activation in \src, and perhaps in other magnetars. They
also point to twists as the source for short bursts, perhaps through
magnetic reconnection in the magnetosphere.

\section*{Acknowledgments}

This research has made use of the \nustar\ Data Analysis Software (NuSTARDAS) 
jointly developed by the ASI Science Data Center (ASDC, Italy) and the 
California Institute of Technology (USA). GY acknowledges support from 
NASA under \nustar\ Guest Observer cycle-4 program 4227, grant number 
80NSSC18K1610. M.G.B. acknowledges the generous support of the 
NSF through grant AST-1813649.

\bibliographystyle{aa}

\end{document}